\documentclass[%
 reprint,
 amsmath,amssymb,
 aps,
]{revtex4-2}

\usepackage{graphicx}
\usepackage{dcolumn}
\usepackage {xcolor}
\usepackage{bm}

\begin{document}

\preprint{APS/123-QED}

\title{Investigation of Gamow-Teller strength of $^{186}$Hg\\within deformed pn-QRPA}

\author{Jameel-Un Nabi}
\altaffiliation{University of Wah, Quaid Avenue, Wah Cantt 47040, Punjab, Pakistan.}

\altaffiliation{Faculty of Engineering Sciences,
	 Ghulam Ishaq Khan Institute of Engineering Sciences and Technology, Topi, 23640, KP, Pakistan}
\email{ jameel@uow.edu.pk}
\author{ Asim Ullah}
\altaffiliation{Faculty of Engineering Sciences, Ghulam Ishaq Khan Institute of Engineering Sciences and Technology, Topi, 23640, KP, Pakistan}
\email{asimullah844@gmail.com}
\author{Abdul Kabir}
\altaffiliation{Department of Space Science, Institute of Space Technology, Islamabad 44000, Pakistan}
\email{kabirkhanak1@gmail.com}
\author{Abdul Muneem}
\altaffiliation{Faculty of Engineering Sciences, Ghulam Ishaq Khan Institute of Engineering Sciences and Technology, Topi, 23640, KP, Pakistan}
\email{abdulmuneemphysics25@gmail.com}
\author{Mahmut B\"{o}y\"{u}kata}
\altaffiliation{Physics Department, Science and Arts Faculty, K\i r\i kkale
	University, 71450, K\i r\i kkale, Turkey}
\email{boyukata@kku.edu.tr}

\date{\today}

\begin{abstract}
	Recently the the total absorption gamma spectroscopy technique was used to determine the Gamow-Teller (GT) distribution of $\beta$-decay of $^{186}$Hg. It was concluded that the best description of the measured data was obtained with dominantly prolate components for both parent $^{186}$Hg and daughter $^{186}$Au. Motivated by the recent findings, we investigate the effect of nuclear deformation on the energy distribution of the GT strength of the decay of $^{186}$Hg into $^{186}$Au within the framework of pn-QRPA based on the deformed Nilsson potential. To do the needful, we first calculate the energy levels and shape prediction of $^{186}$Hg within the interacting boson model. {The computed GT strength distribution satisfied the model independent Ikeda sum rule 100 \% (99.98 \%) for the prolate (oblate) case.} Based on the strength distributions, the deformed pn-QRPA model with separable interaction prefers a prolate shape for the ground-state of $^{186}$Hg and supports the shape coexistence for this nucleus.  
\end{abstract}

\maketitle


\section{\label{Intro}Introduction}
Neutron-deficient Mercury (Hg) isotopes have been studied intensively, both experimentally and theoretically, in the last decades. The motivation behind such investigations is to explore the shape coexistence phenomenon, which is a unique behavior in finite-many-body quantum systems characterized by eigen-states exhibiting {different} shapes in a  {particular atomic}  nucleus~\cite{Hey11,Jul01}. In the nuclide chart such quantum mechanical phenomenon occurs visibly in specified regions~\cite{Hey11,Woo92,Hey83}. The shape coexistence for the Hg isotopes was previously observed from isotope shift measurements \cite{Bon72} in the region $Z$ = 82, which revealed a dramatic transition in the nuclear size between
the ground states of the odd-mass $^{185,187}$Hg isotopes.
Such interesting phenomena are visible in the Hg isotopic chain which is rare in other isotopic chains in the nuclide chart \cite{Ang13}.  The ground states of Hg isotopes (A = 180 - 190) have been discovered to be oblate, with prolate excited states within 1-MeV region \cite{Moreno2006}.

For unstable nuclei, the  decay properties has been studied theoretically with reasonable success. Thus neutron-deficient region of Hg nuclei is considered
as a conventional region for understanding the shape transitions and shape effects. The possible effects of these quantum mechanical shape coexistence have been studied on $\beta$-decay. Therefore, $\beta$-decay can interpret a holistic view relating the nuclear shapes and specifically the exceptional situations \cite{Ham95,Fri95}.

In case of $\beta$-decay of $^{186}$Hg, the ground states around A = 186 of even-even Hg were found well suited with oblate shape which is not so often to be seen in the nuclide chart \cite{Str11}. Meanwhile the odd-A isotopes below A = 186 best match with prolate shapes \cite{Hil90}. A similar phenomenon was also observed for oblate and prolate shape transitions for $^{186}$Au and $^{187}$Au, respectively \cite{Wal87,Wal89,Hin91,Pas94}. This interesting phenomenon shows the importance of $\beta$-decay that may connect partners assumed as dissimilar regarding deformations, particularly in their ground states. These ground states are considered to be, in general, oblate in $^{186}$Hg and prolate in $^{186}$Au in the parent and daughter states, respectively. For these specific cases, decays between partners with similar shapes are argued to be suppressed \cite{Boi15}. In the past, many theoretical approaches have been employed to describe the co-existence phenomenon of many ground states at low energies \cite{Hey11}. The GT strength distributions for the neutron-deficient isotopes around A $\approx$ 70 have been investigated employing the deformed-QRPA theory \cite{Sar11}. Later these studies were extended to \textit{fp}-shell
nuclei \cite{Sar13}. The GT strength distributions for Po, Pb and Hg isotopes were studied in Ref. \cite{Sar05} using a deformed
Skyrme HF+BCS+QRPA approach, which showed the existence of shape isomers and
locality of their equilibrium deformations. By changing the equilibrium deformations, the GT strength distributions displayed specific features
not sensitive to Skyrme and pairing forces. These observed features can be used as trademark of the shape isomers. For GT strength distributions the mean field was expanded to spin-isospin
residual interaction in the well-known proton-neutron (pn)-QRPA model \cite{Mut92,Mol08,Hir91}. The pn-QRPA model has been widely employed for $\beta$-decay
weak rates and half-lives calculations. In pn-QRPA framework the GT residual interactions has two components: particle-hole (ph) and particle-particle (pp) interactions. The \textit{ph} component accounts for the structure and position of GT
resonance \cite{Sar11} while the \textit{pp} component accounts for proton-neutron pairing force in J$^{\pi}$ = 1$^{+}$ coupling channel \cite{Sar11,Hin91,Mut92}. The strength of
these interactions is optimized globally to reproduce the experimental half-lives.
For Z = 6 to 114, the allowed weak decay rates were calculated in the pn-QRPA
framework \cite{Kla84}. Both $\beta$-decay  and electron capture \cite{Hir93} rates were calculated for the nuclei far from stability line.

Recently the $\beta$-decay GT strength distributions of $^{186}$Hg and $^{186}$Au were measured by the total absorption spectroscopy technique {free from the Pandemonium systematic error \cite{Alg21}}. These measured GT strengths were compared with theoretical QRPA
model using SLy4 Skyrme force {\cite{Boi15}}.
The measured GT strength distribution and the half-life were later described by a mixture of oblate and prolate configurations independently in the parent and daughter nuclei. It was concluded that the best description of the experimental beta strength was obtained with dominantly prolate components for both parent $^{186}$Hg and daughter $^{186}$Au nuclei.

The lack of precision on key matrix	elements  prevented the authors from drawing
firm conclusions on the nature of the deformation of the
ground or excited 0$^+$ states of $^{186}$Hg \cite{Bre14}.  
Motivated by this recent development, we re-visit the energy distribution of the GT strength of  $^{186}$Hg  as a function of nuclear deformation and employ the deformed pn-QRPA model, with separable interaction, to compute the GT
strength distributions and half-life of $^{186}$Hg. We use three different values of deformation parameters in our nuclear model corresponding to spherical, oblate and prolate shapes. The spherical deformation of $^{186}$Hg was calculated in the current work using the interacting boson model-1 (IBM-1)~\cite{Iachello87}. In addition an oblate deformation of $^{186}$Hg, obtained using the FRDM approach \cite{Mol16}, and a prolate value, adopted from {the Skyrme interaction based QRPA calcualtion} \cite{Boi15}, was used in our model calculation. The computed GT strength distributions, for the three cases, were compared with the recently reported TAG measurement of the $\beta$-decay of $^{186}$Hg~\cite{Alg21}. 


We present our  {model formalism} in Section~\ref{sec:1}. The  results are discussed in Section~\ref{sec:2}. We finally conclude our investigation in Section~\ref{sec:3}.

\section{Formalism}
\label{sec:1}

The following pn-QRPA Hamiltonian was employed for computing the GT
strength distributions:
\begin{equation} \label{H}
	H^{QRPA} = H^{sp} + V^{pair} + V^{pp}_{GT} + V^{ph}_{GT},
\end{equation}
where $H^{sp}$ stands for the single particle Hamiltonian and $V^{pair}$ is nucleon-nucleon pairing interaction, for which the BCS approximation was considered. $V_{GT}^{pp}$ and $V_{GT}^{ph}$ are the \textit{pp} and \textit{ph} GT interaction potentials, respectively. The wave functions and energies of the single particle were treated in the Nilsson model \cite{nil55} with the incorporation of the nuclear deformation. The oscillator constant was computed using $\hbar\omega=41A^{1/3}$. The Nilsson-potential parameters were chosen from Ref.~\cite{Rag84}.
$Q$-values were calculated from the mass excess values taken from the mass compilation of Audi et al. \cite{Aud17}.
The pairing gaps were calculated using an empirical three term formula based on separation energies of neutron ($S_n$) and proton ($S_p$), given by:
\begin{equation}
	\bigtriangleup_{pp} =\frac{1}{4}(-1)^{Z+1}[S_p(A+1, Z+1)-2S_p(A, Z)+S_p(A-1, Z-1)]
\end{equation}
\begin{equation}
	\bigtriangleup_{nn} =\frac{1}{4}(-1)^{A-Z+1}[S_n(A+1, Z) - 2S_n(A, Z) + S_n(A-1, Z)]
\end{equation}
The spherical basis was transformed to the (axial-symmetric) deformed basis using
\begin{equation}\label{df}
	d^{\dagger}_{m\alpha}=\Sigma_{j}D^{m\alpha}_{j}c^{\dagger}_{jm},
\end{equation}
where $d^{\dagger}$ and $c^{\dagger}$ are particle creation operators in the deformed and spherical basis, respectively. The matrices $D^{m\alpha}_{j}$ were obtained by diagonalizing the Nilsson Hamiltonian.\\ For pairing within BCS approximation, a constant pairing force was applied and a quasi-particle (q.p) basis was introduced:
\begin{equation}\label{qbas}
	a^{\dagger}_{m\alpha}=u_{m\alpha}d^{\dagger}_{m\alpha}-v_{m\alpha}d_{\bar{m}\alpha}
\end{equation}

\begin{equation}
	a^{\dagger}_{\bar{m}\alpha}=u_{m\alpha}d^{\dagger}_{\bar{m}\alpha}+v_{m\alpha}d_{m\alpha},
\end{equation}
where $\bar{m}$, $a^{\dagger}$ and $a$ represents the time reversed state of $m$, the q.p. creation and annihilation operator, respectively which comes in the RPA equation. The occupation amplitudes ($u_{m\alpha}$ and $v_{m\alpha}$) were computed using BCS approximation (satisfying $u^{2}_{m\alpha}$+$v^{2}_{m\alpha}$ = 1).\\
Within the pn-QRPA framework, the GT transitions are described in terms of phonon creation and one describes the QRPA phonons as
\begin{equation}\label{co}
	A^{\dagger}_{\omega}(\mu)=\sum_{pn}[X^{pn}_{\omega}(\mu)a^{\dagger}_{p}a^{\dagger}_{\overline{n}}-Y^{pn}_{\omega}(\mu)a_{n}a_{\overline{p}}],
\end{equation}
where the indices $n$ and $p$ stand for $m_{n}\alpha_{n}$ and $m_{p}\alpha_{p}$, respectively, and differentiating neutron and proton single-particle states. The summation was taken over all proton-neutron pairs satisfying $\mu=m_{p}-m_{n}$ and $\pi_{p}.\pi_{n}$=1, with $\pi$ representing parity. In Eq.~(\ref{co}), $X$ and $Y$ represent the forward- and backward-going amplitudes, respectively, and are the eigenfunctions of the  RPA matrix  equation.  In pn-QRPA theory, the proton-neutron residual interactions work through \textit{ph} and \textit{pp} channels, characterized by interaction constants $\chi$ and $\kappa$, respectively. The $ph$ GT force can be expressed as
\begin{equation}\label{ph}
	V^{ph}= +2\chi\sum^{1}_{\mu= -1}(-1)^{\mu}Y_{\mu}Y^{\dagger}_{-\mu},\\
\end{equation}
with
\begin{equation}\label{y}
	Y_{\mu}= \sum_{j_{p}m_{p}j_{n}m_{n}}<j_{p}m_{p}\mid
	t_- ~\sigma_{\mu}\mid
	j_{n}m_{n}>c^{\dagger}_{j_{p}m_{p}}c_{j_{n}m_{n}},
\end{equation}
and the $pp$ GT force as
\begin{equation}\label{pp}
	V^{pp}= -2\kappa\sum^{1}_{\mu=
		-1}(-1)^{\mu}P^{\dagger}_{\mu}P_{-\mu},
\end{equation}
with
\begin{eqnarray}\label{p}
	P^{\dagger}_{\mu}= \sum_{j_{p}m_{p}j_{n}m_{n}}<j_{n}m_{n}\mid
	(t_- \sigma_{\mu})^{\dagger}\mid
	j_{p}m_{p}>\times \nonumber\\
	(-1)^{l_{n}+j_{n}-m_{n}}c^{\dagger}_{j_{p}m_{p}}c^{\dagger}_{j_{n}-m_{n}},
\end{eqnarray}
where the remaining symbols have their usual meanings.
Here, the different signs in \textit{ph} and \textit{pp} force reveal the opposite nature of these interactions i.e. \textit{pp} force is attractive while the \textit{ph} force is repulsive. The interaction constants $\chi$ and $\kappa$ were chosen in concordance with the suggestion given in Ref.~\cite{hom96}, following a $1/A^{0.7}$ relation. Authors in Ref.~\cite{hom96} performed a systematic study of the $\beta$-decay within the  framework of pn-QRPA and employed a schematic GT residual interaction. The \textit{ph} and \textit{pp} force were consistently included for both $\beta^+$ and $\beta^-$ directions, and their strengths were fixed as smooth functions of mass number A of nuclei in such a way that the calculation best reproduced the
observed $\beta$-decay properties of nuclei. Our computation further fulfilled the model independent Ikeda sum rule~\cite{Ike63}. The reduced transition probabilities for GT transitions from the QRPA ground state
to one-phonon states in the daughter nucleus were obtained as
\begin{equation}
	B_{GT} (\omega) = |\langle \omega, \mu ||t_{+} \sigma_{\mu}||QRPA \rangle|^2
\end{equation}
For further details and complete solution of Eq.~(\ref{H}), we refer to Hirsch et al. \cite{Hir93}.
The partial $\beta$-decay half-lives were calculated using the following relation
\begin{eqnarray}
	t_{p(1/2)} =
	\frac{D}{(g_A/g_V)^2f_A(Z, A, E)B_{GT}(\omega)},
\end{eqnarray}
where $\omega$ is the final state energy, $E$ = $Q_{EC}$ - $\omega$, $g_A/g_V$ (= -1.254)\cite{war94} represents ratio of axial vector to the vector coupling constant and D = $\frac{2\pi^3 \hbar^7 ln2}{g^2_V m^5_ec^4} = 6295 s$. {Phase space factor $f_A(Z, A, E)$ is the Fermi integral for axial vector transitions}. $B_{GT}$  are the reduced transition probabilities for the GT  transitions. The $\beta$-decay half-life of $^{186}$Hg was determined by summing up all transition probabilities to states in the daughter nucleus with excitation energies lying within the $Q_{EC}$ window:
\begin{equation}
	T_{1/2} = \left(\sum_{0 \le E_j \le Q_{EC}} \frac{1}{t_{p(1/2)}}\right)^{-1}
\end{equation}
\section{Results and Discussion}
\label{sec:2}
\begin{figure}[h!]
	\includegraphics[width=0.55\textwidth]{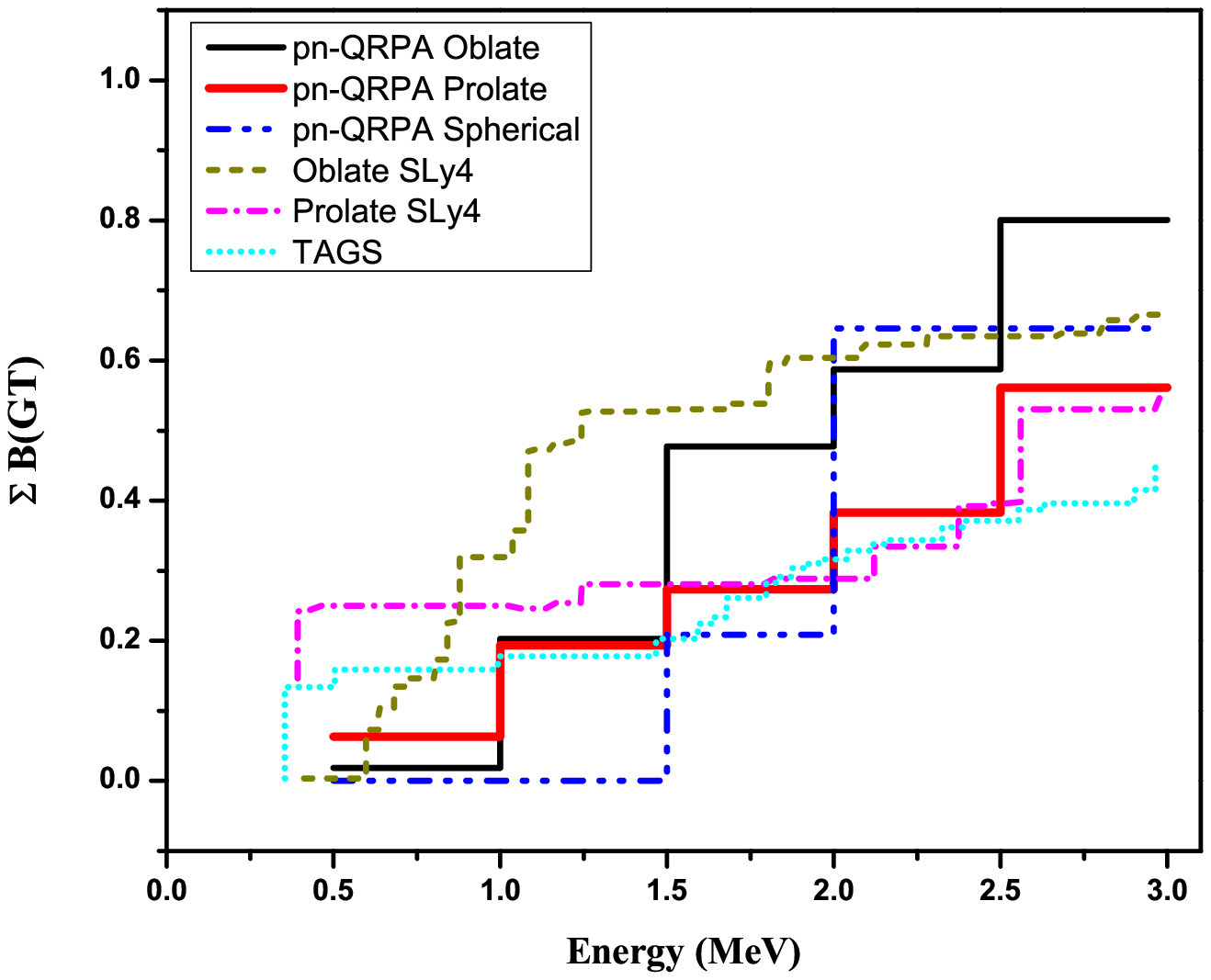}
	\caption{{Comparison of pn-QRPA computed GT strength distributions for
		$^{186}$ Hg decay with the measured \cite{Alg21} and SLy4 calculation \cite{Boi15}}}
	\label{fig:0}
\end{figure}

The current study examines the GT strength and half-life of $^{186}$Hg based on the nuclear shapes.
The quadrupole deformation parameter ($\beta_2$) for the $^{186}$Hg was investigated to demonstrate
the importance of deformation in isotopic evolution. {The IBM-1 model deduced a spherical shape for $^{186}$Hg. An oblate minimum with  $\beta=-0.146$ was reported for  $^{186}$Hg in Ref.~\cite{Mol16}. Moreover Ref.~\cite{Boi15} reported a prolate shape with $\beta=+0.26$ for the same nucleus. These three values of deformation parameter were used to calculate the GT strength distributions and half-lives of $^{186}$Hg within the pn-QRPA model}. \\
The results for $\beta$-decay half-lives of $^{186}$Hg as a function of deformation parameters are shown in Table~\ref{Parameters}. Our computed half-lives of $^{186}$Hg are compared with the self-consistent deformed
Skyrme Hartree-Fock mean field calculation of Ref.~\cite{Moreno2006}. It is to be noted that the latter model did not use a spherical $^{186}$Hg for half-life calculation. It is evident from Table~\ref{Parameters} that neither form has a distinct advantage over the other.
In Fig.~\ref{fig:0} we present our computed GT strength {distributions (using the three deformation parameters). Shown also are the SLy4 oblate and prolate results calculated within the QRPA model~\cite{Boi15}. The recently maesured TAG results (upper limit) are also displayed~\cite{Alg21}. The cumulative GT strength distributions are plotted up to 3~\textit{MeV} excitation energy. Our calculated results show a good comparison with the TAG results till the mid-range energy values for prolate and oblate shapes of $^{186}$Hg nucleus. It is to be noted that for the computation of GT strength distributions, we did not incorporate any explicit quenching factor and hence our computed GT strength is bigger than the experimental and SLy4 data. From Fig.~\ref{fig:0}, one can notice that the oblate shape results in a bigger GT strength at low excitation energies (also evident from Table~\ref{SD}). The sensitivity of the GT strength distributions to the nuclear deformation can produce insight about the nuclear shape in the neutron-deficient Hg nucleus.}\\
\begin{table*}
	\centering\scriptsize\caption{Comparison of calculated half-lives  with the results of Ref.~\cite{Moreno2006}, with different assumed geometric shapes, for $^{186}$Hg. } \label{Parameters}
	\begin{tabular}{|c|c|c|}
		\hline
		Shape &  {$T_{1/2}$ (This work)}& $T_{1/2}$ \cite{Moreno2006}  \\	
		&  $sec$ & $sec$  \\	
		\hline
		Spherical&  78.9 & -- \\
		Oblate  &  54.7 & 47.2 \\
		Prolate &  {75.7} & 68.4  \\
		\hline
		
	\end{tabular}
\end{table*}
\begin{table*}
	\centering\scriptsize\caption{Comparison of experimental $Q_{EC}$ value, measured~\cite{Alg21} and calculated half-lives, total $B(GT_+$) within the EC window, total $B(GT_+$) and $B(GT_-$) within the whole energy range and \% of ISR fulfilled by the current pn-QRPA model and those of Ref.~\cite{Moreno2006} for  $^{186}$Hg. It is to be noted that the pn-QRPA cut-off value for daughter excitation energy ($E_x <$ 60 MeV) is double that considered by Ref.~\cite{Moreno2006} ($E_x <$ 30 MeV). } \label{SD}
	\begin{tabular}{|c|c|c|c|c|c|c|c|c|}
		\hline
		Model& $\beta_2$ & $Q_{EC}$ \cite{Aud17}&  $T_{1/2}$ (Exp) & $T_{1/2}$ (Theory) &$\sum_{Q_{EC}} B(GT_+)$ &$\sum B(GT_+)$& $\sum B(GT_-)$ &\% ISR\\	
		\hline
		& 0.000 (This work)        &       &      & 78.9 &1.29 &13.45& 91.05 &99.48\\
		pn-QRPA  &  -0.146 \cite{Mol16}     & 3.176 & 82.3 & 54.7 &1.15 &16.02& 94.03 &99.98\\
		(This work)  & {0.260} \cite{Boi15}       &       &      & {75.7} &{0.99} &{14.39}& {92.39} &{100.0}\\
		\hline
		pn-QRPA \cite{Moreno2006}&Oblate    & 3.176 & 82.3 & 47.2 &1.22 &2.57& 79.85 &99.08\\
		& Prolate   &       &      & 68.4 &0.89 &2.46& 79.76 &99.10\\
		\hline
	\end{tabular}
\end{table*}
Table~\ref{SD} shows the experimental data ($Q_{EC}$ \cite{Aud17} and $T_{1/2}$ \cite{Alg21}) and our computed results ($T_{1/2}$, total $B(GT_+$) within EC window, total $B(GT_+$) and $B(GT_-$) within the whole energy range and \% of ISR fulfilled by our results). We also compare our calculations with those performed by Ref.~\cite{Moreno2006} in~Table \ref{SD}. It is to be noted that Ref.~\cite{Moreno2006} incorporated a quenching factor of 0.59 for the ratio of axial vector to vector coupling constants. Their cut-off value for daughter excitation energy was 30~$MeV$ which was half of what we used in our calculation. These factors may explain the smaller total GT strengths calculated by Ref.~\cite{Moreno2006}. {It is to be noted that our computed GT strength distribution completely obey the model independent Ikeda sum rule only for the prolate shape. The percentage fulfillment for spherical and oblate cases are 99.48 \% and 99.98 \%, respectively.}

\section{Conclusions}
\label{sec:3}
In this study, we investigated the GT strength  distributions in neutron-deficient $^{186}$Hg nucleus by employing the deformed pn-QRPA framework.  The residual interaction constants $\chi$ and $\kappa$ were adopted from Ref.~\cite{hom96} which reproduced the experimental half-life with a deviation of around 4\%, 34\% and {8\%} for the spherical, oblate and prolate shapes, respectively. We found that the GT strength distributions computed, assuming spherical, prolate and oblate shapes, were characterized by strong peaks at high excitation
energies involved in the $\beta^+$ decay and are more sensitive to the nuclear shapes. {Our computed GT strength distribution completely fulfilled the model independent Ikeda sum rule \textit{only} for the prolate shape}.

 Our current investigation supports the findings of \cite{Alg21} that a dominant prolate component for both  $^{186}$Hg and $^{186}$Au best describes the recently reported measured data.

\section*{Acknowledgements}
{
	M. B\"{o}y\"{u}kata would like to acknowledge the support of Higher
	Education Council of Turkey through project number MEV 2019-1745 of project-based international exchange program.
}



\end{document}